\documentclass[preprint]{article}
\usepackage{graphicx}

\newcommand{\be}{\begin{equation}}
\newcommand{\ee}{\end{equation}}
\newcommand{\ba}{\begin{eqnarray}}
\newcommand{\ea}{\end{eqnarray}}

\begin{document}
\setlength{\textheight}{8.5 in}
\begin{titlepage}

\vskip 5mm

\begin{center}
{\Large\bf {Dimensional Reduction by Conformal Bootstrap }}
\vskip 6mm
 Shinobu Hikami 

\vskip 5mm

Mathematical and Theoretical Physics Unit,
Okinawa Institute of Science and Technology Graduate University,
Okinawa, Onna 904-0495, Japan; hikami@oist.jp

\end{center}
\vskip 20mm
\begin{center}
{\bf Abstract}
\end{center}
\vskip 3mm
The dimensional reductions in  the branched polymer and the random field Ising model (RFIM) are discussed by a conformal bootstrap method. The small size minors are applied for the evaluations of  the scale dimensions of these two models and the results are compared to $D'=D-2$ dimensional Yang-Lee edge singularity and to pure $D'=D-2$ dimensional Ising model, respectively.  For the  former case, the dimensional reduction is shown to be valid for $3 \le D \le 8$, and  for the  later case,  the deviation from the dimensional reduction can be seen below five dimensions.
 \end{titlepage}

\newpage
\section{Introduction}
\vskip 2mm

   The critical exponent of $D$ dimensional branched polymer, which is a  polymer with trivalent
  branches in a $D$ dimensional solvent, is known to be same as  the critical exponent of $D'$ dimensional Yang-Lee edge singularity with $D'=D-2$. The dimension $D$ has a range of $3 \le D \le 8$ for this correspondence. This remarkable correspondence has been explained by the supersymmetry \cite{Parisi1981} for such range of the dimension.  There is also perturbational analysis near 8 dimensions by 
  $\epsilon$ expansion, which supports this dimensional reduction. This dimensional reduction has been proved rigorously for the branched polymer \cite{Brydges2003,Cardy2001,Cardy2003}. 
  
  A random magnetic field Ising model (RFIM) was conjectured to have the dimensional reduction to pure Ising model without magnetic field by the supersymmetric formulation; namely the critical exponents of RFIM in $D$ dimension are same as the critical exponents of the pure Ising model in $D'=D-2$ dimension for $3 < D \le 6$ \cite{Parisi1979}. This conjectured was however
  disproved by the counterexample of $D=3$. It has been shown that there is a phase transition in $D=3$ for RFIM \cite{Imbrie1984}.
    
    The failure of the supersymmetric argument for the dimensional reduction of RFIM  is related to the negative sign of the measure of the functional integral after the integration of  the Grassmann fields.   The instability of fixed point  in $\epsilon$ expansion has been also discussed \cite{Brezin1998}. The formation of a bound state  has been proposed \cite{Brezin2000,Parisi2002}.  The problem of the dimensional reduction of RFIM  remains  for  forty years. The  review article \cite{Sourlas2017} provides the recent references of RFIM including a numerical analysis. We will investigate  RFIM in this article by the conformal bootstrap method, which has been applied to the Yang-Lee edge singularity \cite{gliozzi2013,gliozzi2014,Hikami2017a} and for the branched polymer \cite{Hikami2017b}. We will  discuss again  the branched polymer case  to see a clear difference to RFIM by the conformal bootstrap method.
  
 The conformal bootstrap method  was developed long time ago \cite{Ferrara1973}, and it was applied to critical phenomena \cite{Parisi1971,Symanzik1972} as  an approximation.
 The modern numerical approach was initiated by \cite{Rattazzi2008}. 
The recent studies by this conformal bootstrap method led to many remarkable results  for various symmetries, which references may be  found in a recent review article \cite{Rychkov2018}.

The result of the conformal bootstrap method, in this article is consistent of  the dimensional reductions in the case of branched polymer and RFIM. Since our analysis is limited to  the small size of the determinant of the conformal block, the result should be interpreted as  an approximation for the critical exponent. This method, however as shown in Yang-Lee edge singularity \cite{Hikami2017a}, has an advantage to standard $\epsilon$ expansion, since it estimates  critical exponents in a wide region of the space dimensions, not restricted to the area  near the upper critical dimensions.

For the RMFI, the result suggests the dimensional reduction will hold near $D=6$ up to $D=5$, but below $D=5$, the values of the critical exponent of $D$ dimensions deviate from the conjecture that the dimensional reduction holds; corresponding critical exponents in $D'=D-2$ dimension are the critical exponents of  Ising model.
 
We call the method, which we employ  in this paper, as a determinant method simply in this article.
With  the restriction to small numbers of the operators, the determinant method  has been applied successfully on Ising model, and Yang-Lee edge singularity\cite{gliozzi2013,gliozzi2014,Hikami2017a}. This determinant method can be  applied to the non-unitary cases, since it does not require the unitarity bound. The unitary case, such as $O(N)$ vector model ($N \ge 1$), shows a kink in the boundary curve of the unitarity bound. If we identify this kink as a critical point, we obtain the value of the critical exponent.  Yang-Lee edge singularity and $O(N)$ vector model for $N < 1$ are however not unitary, since the operator coefficients become negative. The disordered systems, like branched polymer and RFIM are described by non-unitary model. The determinant method (or truncation method) may be 
useful for obtaining the critical exponents for the non-unitary models \cite{Rychkov2018}. At the moment we don't know other method for the non-unitary cases.

The branched polymer and RFIM are expressed by the replica limit $N\to 0$ of the $N$ component Ginzburg-Landau effective action. 
We use a few of scale dimensions $\Delta$ for the analysis for bootstrap method instead of the infinite numbers of different $\Delta$. We consider only small value of spin $L$, namely $L=2$ and $L=4$. For the disordered system, the degeneracy of two different scalar $\Delta$ becomes essentially important.
In this paper, we introduce  one scalar scaling dimension $\Delta_1$, which is chosen as a free parameter, in addition to the basic energy scale dimension $\Delta_\epsilon$. It is known that $\Delta_\epsilon$ is a scale dimension for the energy density, and related to the critical exponent $\nu$. The other scaling dimension $\Delta_1$ appears as the traceless symmetric tensor scaling dimension in O(N) vector model. For polymer case, expressed in  the replica limit $N=0$ in O(N) vector model, this scaling dimension $\Delta_1$ coincides with the energy density scale dimension $\Delta_\epsilon$ \cite{Hikami2017b,Shimada2016}.
 In the single polymer and the branched polymer cases, this $\Delta_1$ is same as $\Delta_T = D - \hat\varphi/\nu$, where
$\hat\varphi$ is a crossover exponent of $O(N)$ vector model \cite{Hikami1974}. Thus the degeneracy of two scalar $\Delta$ occurs as $\Delta_T=\Delta_\epsilon$. In RFIM, we do not necessarily assume that $\Delta_1$ is same as $\Delta_T$, as expected. Indeed, the effective replica Hamiltonian is different from polymers as we will see in (\ref{effective}). We assume the value of $\Delta_1$ to be near the value of $\Delta_\epsilon$ similar to the polymer case. Although  $\Delta_1$ is assumed to be different from the scale dimension  $\Delta_\epsilon' = D + \omega$, where $\omega$ is an exponent of the correction to scaling, since $\Delta_1$ and $\Delta'_\epsilon$ are both scalar scale dimensions. 
There is no reason that $\Delta_1$ is same as the scale dimension $\Delta_\epsilon$, when  the dimensional reduction due to the supersymmetry does not hold for RFIM. 

This paper is organized as follows:  In section 2,  Yang-Lee edge singularity is shortly reviewed as an example of the
application of the determinant method. In section 3, a brief review of the conjecture of the random magnetic field Ising model (RFIM)  by the renormalization group and by the supersymmetric argument, which leads to the conclusion that RFIM is equivalent to D-2 dimensional pure Ising model. In section 4, the dimensional reduction of  the branched polymer to Yang-Lee edge singularity is explained by the supersymmetric argument similar to RFIM.  In section 5, we discuss the dimensional reduction of the branched polymer to Yang-Lee edge singularity by the  determinant method. In the section 6, we discuss  RFIM by the determinant method, and see the validity of the dimensional reduction to the pure Ising model.
The section 7 is devoted to summary and discussions. The explanation of the determinant method has been presented in the previous articles  in \cite{Hikami2017a,Hikami2017b}. The various related  notations are also represented in them. We do not repeat these fundamental equations for the brevity, and recommend to consult these equations in the previous articles \cite{Hikami2017a,Hikami2017b}.
\vskip 2mm
\vskip 3mm
\section{Yang-Lee singularity in $1 \le D \le 6$ }
\vskip 2mm
We first consider  the Yang-lee edge singularity \cite{Fisher1978}, since the D dimensional branched polymer has a dimensional reduction to D-2 dimensional Yang-Lee edge singularity. 
 Yang-Lee edge singularity is a good example  of the determinant method, which we will apply on RFIM later.
It is originated from the critical behavior of the density of the zeros of the partition function of Ising model with a complex magnetic field. It is described by $\phi^3$ field theory
with an imaginary coupling constant. This Yang-Lee edge singularity has been studied by the conformal bootstrap method \cite{gliozzi2013,gliozzi2014,Hikami2017a}.

We consider a finite scale dimensions $\Delta$. The four scaling dimensions, $\Delta_\phi$,  $\Delta_\epsilon$, $\Delta_\epsilon'$ denote the scaling dimensions for a  field $\phi$, an energy $\epsilon$ and a correction to scaling scaling $\omega$, respectively and we include also $Q$ (a  spin 4 operator, fourth derivatives).   The definition of these four parameters can be found in \cite{gliozzi2013} and \cite{Hikami2017a}. For Yang-Lee edge singularity, which is described by $\phi^3$ theory with an imaginary coefficient, the constraint of the degeneracy due to the equation of motion,  $\Delta_\phi= \Delta_\epsilon$ is imposed. In the map of zero loci of $4\times 4$ minors, there appear several intersection points of zero loci lines. In a previous article \cite{Hikami2017a}, we discussed the
reason for the existence of such intersection points of three or more lines of the zeros of minors by the Pl\"ucker relations.  Below we repeat the results  of D=6, 4 and 3 for Yang-Lee edge singularity, which has been investigated in \cite{Hikami2017a}.

\vskip 2mm
\noindent{\bf{D=6}}
\vskip 2mm
In Fig.1, the  zero loci of $4\times 4$  minors in $D=6$ intersect at three fixed points with parameters of $Q=8$ and $\Delta_\epsilon'=5.9$.
 The upper one is a free field fixed point with $\Delta_\phi=2.0$ and $\Delta_\epsilon=4.0$, and the middle intersection point is the continuation of the non trivial fixed point of Wilson-Fisher to six dimensions (infrared unstable).  The lower fixed point  corresponds to Yang-Lee edge singularity ($\Delta_\epsilon = \Delta_\phi= 2.0$). The horizontal line at $\Delta_\epsilon = 2.0$ shows a pole of $\Delta= \frac{D-2}{2}$ for $D=6$ \cite{Kos2015,Kos2016}.

 \vskip 2mm
 \noindent{\bf{D=4}}
 \vskip 2mm
 In Fig.2, with D=4, $Q=6.0$ and $\Delta_{\epsilon'}=4.0$, the intersection points appear at (i) $\Delta_\phi=1.0$, $\Delta_\epsilon= 2.0$, which is Wilson-Fisher free field point , and (ii) Yang-Lee fixed point,
 which is located at $\Delta_\phi= 0.929123, \Delta_\epsilon =  0.922221$. The  values of (ii) can be compared to $\Delta_\phi=0.83175$. To obtain better value, $Q$ is chosen as
 $Q=5.712$, then the intersection point moves to $\Delta_\phi =0.827562,\Delta_\epsilon=0.871742$. This is close to the result by Pad\'e analysis , which gives $\Delta_\phi=0.83175$.
\begin{figure}
{\bf Fig. 1}
\hskip 20mm {$\Delta_\epsilon$}\\
\centerline{\includegraphics[width=0.5\textwidth]{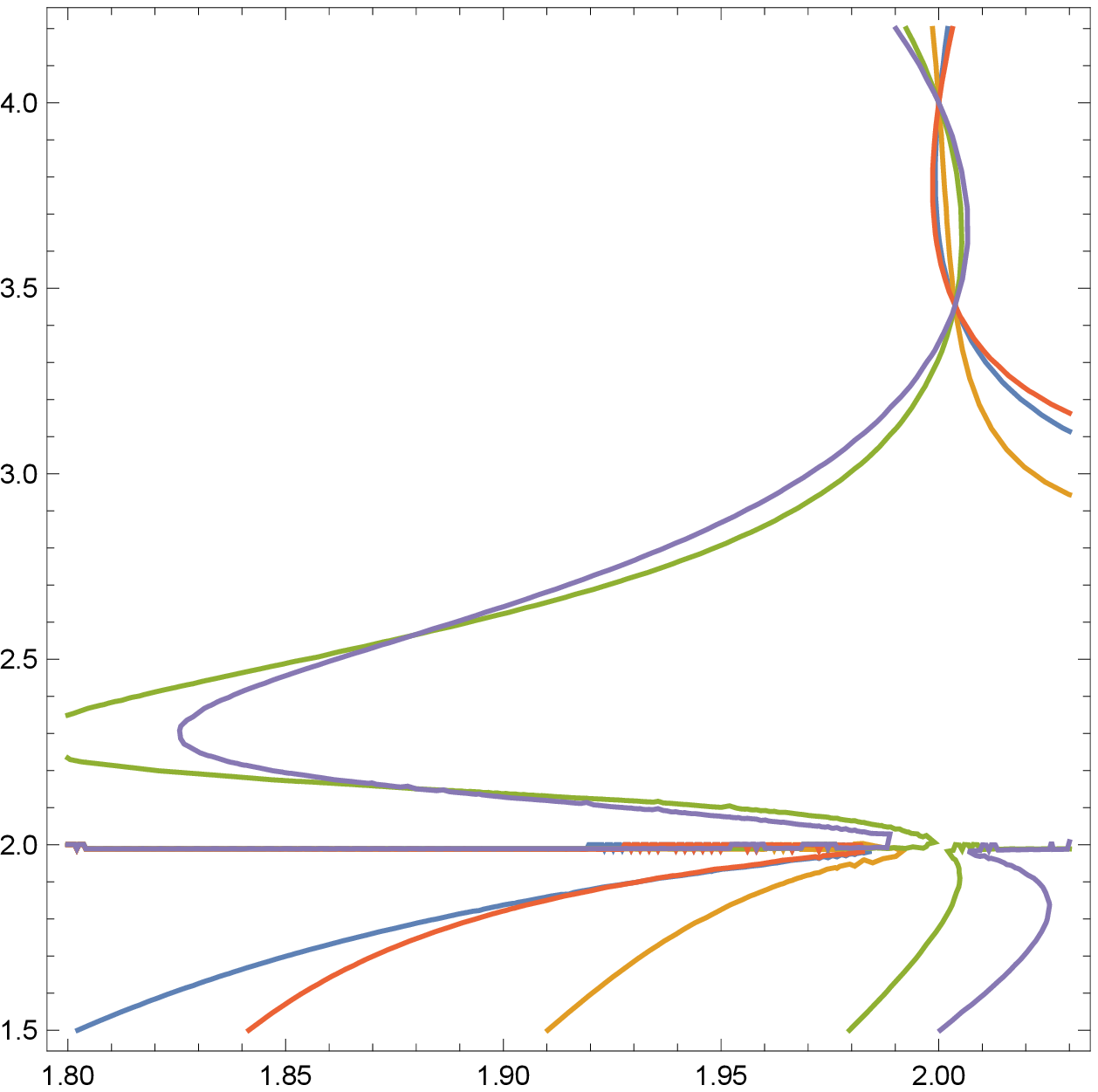}}\\
\hskip 90mm {$\Delta_\phi$}\\

\caption{D=6 (Ising and Yang-Lee) : The  zero loci of $4\times 4$  minors  intersect at three fixed points, The upper one is free field fixed point$(\Delta_\phi,\Delta_\epsilon) = (2.0,4.0)$, and the middle is the continuation of the non trivial fixed of Wilson-Fisher to six dimensions (infrared unstable).  The lower fixed point  corresponds to Yang-Lee edge singularity ($\Delta_\epsilon = \Delta_\phi=2.0$). }
\end{figure}

\begin{figure}
{\bf Fig.2}\hskip 20mm {$\Delta_\epsilon$}\\
\centerline{\includegraphics[width=0.5\textwidth]{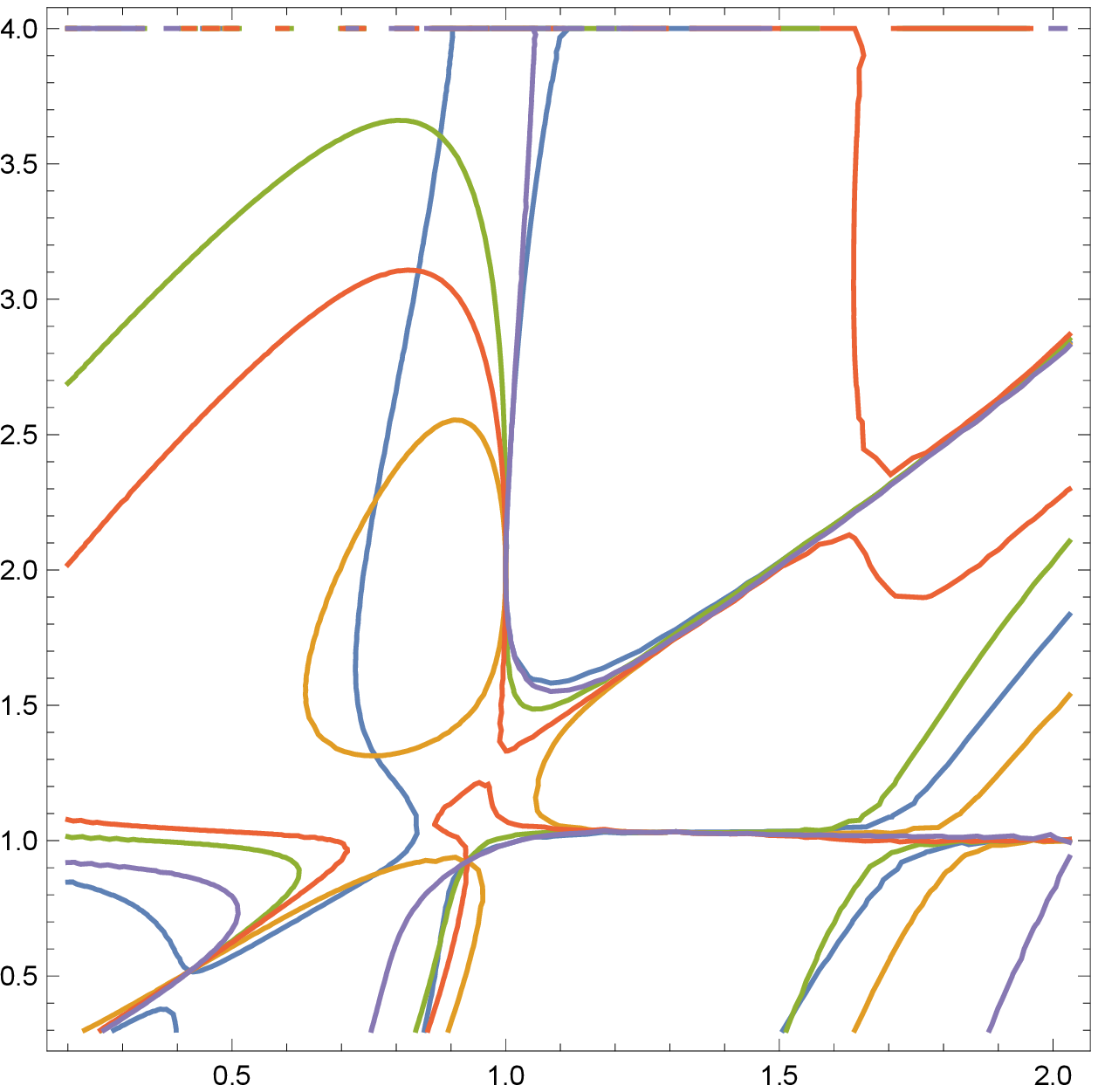}}\\
\hskip 90mm { $\Delta_\phi$}\\

\caption{Yang-Lee in D=4: The  zero loci of $4\times 4$  minors  intersect at three fixed points. The  free field fixed point for $\phi^4$ theory at $\Delta_\phi=1,\Delta_\epsilon=2$ is shown as an intersection point of the 5 minor loci.  The lower fixed point  corresponds to Yang-Lee edge singularity ($\Delta_\epsilon = \Delta_\phi=0.8$). The parameters  $Q=6.0$ and $\Delta_{\epsilon'}=4.0$ are chosen.
}
\end{figure}

\vskip 2mm
\noindent
{\bf { D = 3}}
\vskip 2mm

The intersection map of $4\times 4$ minors $d_{ijkl}$ depend upon the parameters of $Q$ and $\Delta_\epsilon'$. There are Ising model fixed point and Yang-Lee fixed point,
but their parameter $Q$ and $\Delta_\epsilon'$ are different.
When the parameters $Q=4.75,\Delta_\epsilon'=5.0$ are chosen, the Yang-Lee intersection point becomes $\Delta_\phi=0.2314,\Delta_\epsilon=0.2316$. 
For this parameters, the intersection point of Ising model disappears because the parameter $Q$ is far from the correct value ($Q=5.02$) of Ising model.
From the Pad\'e analysis, the scale dimension $\Delta_\phi=\Delta_\epsilon$ is obtained as $\Delta_\phi= 0.2299$ \cite{Hikami2017a}.

The values of $\Delta_\phi=\Delta_\epsilon$ , which are obtained by determinant method in \cite{Hikami2017a} are listed in general dimensions in Table 1, which will be used in  the discussion of the dimensional reduction.
\vskip 50mm
{\bf Table 1 : The scale dimensions of Yang-Lee model \cite{Hikami2017a}} (* exact value).\\
\begin{tabular}[t]{|c|c|c|c|}
\hline
& $\Delta_\phi=\Delta_\epsilon $& Q& $\Delta_\phi$  (Pad\'e)\\ 
\hline
D=2.0 & - 0.4* & 3.6* & - \\
D=3.0 & 0.174343  & 4.34106   & 0.22995\\
D=3.5 & 0.499401 & 5.04195 & 0.53153 \\
D=4.0 & 0.823283 & 5.71152 & 0.83175 \\
D=4.5 & 1.13755  & 6.33395   & 1.1300 \\
D=5.0 & 1.43807 & 6.91716  & 1.4255 \\
D=5.5 & 1.72469 & 7.46985   & 1.7165 \\
D=6.0 & 2.0 & 8.0  & 2.0\\
\hline
\end{tabular}

\newpage
 
\section{Random magnetic field Ising model (RFIM)}
\vskip 2mm
The dimensional reduction in the D dimensional random magnetic field Ising model to the pure Ising model in D-2 dimension has been discussed intensively 
, by a diagrammatic perturbation \cite{Imry1975} and by a supersymmetric argument \cite{Parisi1979}. Their results support the dimensional reduction to pure
Ising model in $D-2$ dimensions.
However, this dimensional reduction for RFIM was found to be incorrect. Particularly for the lower dimension, it has been proved that the lower dimension is not three \cite{Imbrie1984}. 

There are several suggestions  for the reason of this breakdown. 
It is recognized that RFIM is related to the replica symmetry breaking like a spin glass problem due to negative sign, since the measure expressed as determinant by Grassmannian variables,  can be negative. Also it was shown that the fixed point becomes unstable with introducing  more relevant couplings.
\cite{Brezin1998,Brezin2000}.

This puzzling problem stands for a  long time from the beginning of the renormalization group study more than forty years. It is known that the dimensional reduction works for (i) the branched polymer in D dimension, which is equivalent to Yang-Lee edge singularity in $D'=D-2$ dimensions, (ii) electron density of state in  two dimensional  random impurity potential  under a strong magnetic field.
\cite{Brezin1984,Wegner2015}. 

We  briefly summarize in the following the argument of the dimensional reduction of RFIM.
The  application of the replica method to RFIM  is the replacement of the following action,
\be
S(\phi) = \int d^Dx [\frac{1}{2}(\nabla \phi)^2 + \frac{1}{2}r \phi^2 + \frac{1}{8}g \phi^4 - h(x)\phi(x)]
\ee
by
\be\label{effective}
S(\phi_\alpha) = \int d^D x [\sum_{\alpha=1}^N( \frac{1}{2}(\nabla \phi_\alpha)^2 + \frac{1}{2} t \phi_\alpha^2 + \frac{1}{8} u  \phi_\alpha^4) - \frac{c}{2}\sum_{\alpha,\beta=1}^N
\phi_\alpha \phi_\beta]
\ee
with a random magnetic field $h(x)$, which obeys the white noise distribution,
\be
<h(x)> =0,\hskip 3mm <h(x)h(x')> = c \delta(x - x'). 
\ee
The quenched average requires about $W={\rm log} Z$, and the replica $N\to 0$ limit takes this average about $Z$ as
\be
\lim_{N\to 0} \frac{1}{N}( < Z^N > -1) = < {\rm log}  Z>.
\ee 
Under this $c$, propagator $G$ in the replica follows as
 \be\label{G}
 G^{\alpha,\beta}(q) = \frac{\delta_{\alpha\beta}}{q^2 + t} + \frac{c}{(q^2+ t)(q^2 +  t - N c)}
 \ee
 The loop expansion of this propagator shows a critical dimension at $D=6$ in the limit $N\to 0$ since the propagator changes due to the non-vanishing $c$, like as
 \be
 I(p) = \int d^D q \frac{1}{(q^2+ t)(q^2+t- N c)((p-q)^2+t)((p-q)^2+ t-N c)}
 \ee
 Indeed $\epsilon=(6-D)$ expansion gives
 \be\label{ON}
 \frac{1}{\nu} = 2 - \frac{N+2}{N+8}\epsilon + O(\epsilon^2).
 \ee
 which has a same form of ordinary O(N) vector model without a random magnetic field of $\epsilon = 4-D$.
 
 Parisi and Sourlas \cite{Parisi1979} introduced anti-commuting variables instead of the replica field, which plays with - 2 dimensions, in the stochastic field formulation.
 
 The Green function $G(x) =<\phi(x)\phi(0)>$ is 
 \ba
 G(x) &=& \int D\phi D\omega D\psi \phi(x)\phi(0) {\rm exp}[- \int d^D y (-\frac{1}{2} \omega^2 + \omega[-\Delta \phi + V'(\phi)]\nonumber\\
 && + \bar \psi[-\Delta + V^{''}(\phi)]\psi)]\nonumber\\
 &=& \int D\phi D h \phi(x)\phi(0) \delta(-\Delta \phi + V'(\phi) + h) {\rm det}[-\Delta + V^{''}(\phi)] \nonumber\\
 && \times {\rm exp}[-\frac{1}{2}\int h^2(y) d^Dy]
 \ea
 where $V(\phi) =\frac{1}{2} m^2 \phi^2 + g\phi^4$. The short coming of this formulation is the sign of the determinant, which can be negative. There appears a supersymmetric
 BRST gauge transformation \cite{Neveu1986},
 \ba
&& \delta \phi= - \bar a \epsilon_\mu x_\mu \psi, \hskip 2mm
 \delta \omega = 2 \bar a \epsilon_\mu \partial_\mu \psi,\nonumber\\
 &&\delta \psi=0,\hskip 2mm \delta \bar \psi = \bar a (\epsilon_\mu x_\mu \omega + 2 \epsilon_\mu \partial_\mu \phi)
 \ea
 where $\bar a$ is an infinitesimal anticommuting number, $\epsilon_\mu$ is an arbitrary vector.
 With the superfield $\Phi(x,\theta)$,
 \be
 \Phi(x,\theta) = \phi(x) + \bar \theta \psi(x) + \bar\psi(x) \theta + \theta \bar\theta \omega(x)
 \ee
 the Lagrangian becomes 
 \be
 L(\Phi) = -\frac{1}{2}\Phi \Delta_{ss} \Phi + V(\Phi)
 \ee
 with $\Delta_{ss}= \Delta+ \partial^2/\partial \bar\theta \partial \theta$. The superspace $(x,\theta)$ is equivalent to D-2 dimensional space. Therefore anticommuting coordinate has a negative dimension - 2. This may give a possible proof of the dimensional reduction from D to D-2, but as we discussed before, this dimensional reduction does not work, since the
 measure does not show the positivity.

 \vskip 2mm
 \section{Branched polymer }
\vskip 2mm

We briefly consider the branched polymer since the formulation is very close to RFIM. The main difference is that the effective Hamiltonian is $\phi^3$ instead of $\phi^4$ for the branched polymer.
This makes the upper critical dimension as eight for the branched polymer. (RFIM has the upper critical dimension as six). The branched polymer is described by the branching terms in addition to the self-avoiding term.
We write the action for a $p$-th branched polymer as $N$-replicated field theory
 \be
S = \int d^D x \bigg(\frac{1}{2}\sum_{\alpha=1}^N ( (\nabla \phi_\alpha)^2 - \sum_{p=1}^\infty u_p \phi_\alpha^p ) + \lambda (\sum_{\alpha=1}^N \phi_\alpha^2)^2 \bigg) 
 \ee
 
 The term $\phi_\alpha^p$ represents the $p$-th branched polymer. After the  rescaling and neglecting irrelevant terms, the following action is obtained
 
  \be\label{branch}
 S = \int d^D x \bigg(\frac{1}{2}\sum_{\alpha=1}^N ((\nabla \phi_\alpha)^2 + V(\phi_\alpha)) +  C\sum_{\alpha,\beta=1}^N
 \phi_\alpha \phi_\beta \bigg) 
  \ee
with $V(\phi_\alpha)=t \phi_\alpha -\frac{1}{3}\phi_\alpha^3 + O(\phi_\alpha^4)$.

In the paper of Parisi-Sourlas \cite{Parisi1981},  the equivalence to Yang-Lee edge singularity was shown by the supersymmetric argument.  
The $\epsilon$ expansion of the critical exponent $\eta$ of the branched polymer was studied \cite{Fisher1978,Lubensky1979},

\be
\eta = - \frac{1}{9}\epsilon
\ee
where $\epsilon= 8 - D$.
The scaling dimension $\Delta_\phi$ becomes
\be
\Delta_\phi = \frac{D - 2 + \eta}{2} 
\ee
In this formula, we put $D \to D-2$, and $\epsilon \to \epsilon = 6 -D$, then we get
\be
\Delta_\phi = 2 - \frac{5}{9}\epsilon
\ee
where $\epsilon = 6-D$.
This last formula is exactly same as the expansion of Yang-Lee edge singularity, $\Delta_\phi = 2 - \frac{5}{9} \epsilon$, with $\epsilon = 6 - D$.

The exponent $\nu$ of Yang-Lee edge singularity $(\epsilon= 6 - D$) is
\be
\frac{1}{\nu}= \frac{1}{2}(D+ 2 - \eta) = \frac{1}{2}(8 - \epsilon + \frac{1}{9}\epsilon) = 4 - \frac{4}{9}\epsilon
\ee

This reads up to order $\epsilon$,
\be
\Delta_\epsilon = D - \frac{1}{\nu} = (6-\epsilon) - (4 -\frac{4}{9}\epsilon) = 2- \frac{5}{9}\epsilon = \Delta_\phi
\ee

Thus the values of exponents $\eta$ and $\nu$ of the branched polymer become same as the exponents of Yang-Lee edge singularity. This holds all orders of $\epsilon$ due to  the
equation of motion. 
The scale dimensions of $\Delta_\epsilon$ and $\Delta_\phi$, however become different since they involve the space dimension $D$ explicitly.
For instance, in the branched polymer at $D=8$,
\be\label{equivalence}
\Delta_\epsilon = 8 - \frac{1}{\nu}=4, \hskip 3mm
\Delta_\phi = 3
\ee
where
for Yang-Lee edge singularity at D=6,
\be
\Delta_\epsilon= 2,\hskip 3mm \Delta_\phi = 2.
\ee

In general dimension $D \le 8$, from the relation  to Yang-Lee edge singularity, we have for the branched polymer,
\be\label{super}
\Delta_\epsilon = \Delta_\phi + 1
\ee
as shown in (\ref{equivalence}) for $D=8$. 
We get the following relations by noting the difference of the dimension $D$ for two cases,
 \ba\label{brYL1}
&& \Delta_\phi ({\rm branched\hskip 1mm polymer \hskip 1mm in \hskip 1mm D \hskip 1mm dim.}) = \Delta_\phi ({\rm YangLee \hskip 1mm in \hskip 1mm D' \hskip 1mm dim.}) + 1,\nonumber\\
&& \Delta_\epsilon ({\rm branched\hskip 1mm polymer \hskip 1mm in \hskip 1mm D \hskip 1mm dim.}) = \Delta_\epsilon ({\rm YangLee \hskip 1mm in \hskip 1mm D' \hskip 1mm dim.}) + 2.
 \ea
 where $D' = D-2$.
This relation is related to the ${\mathcal N}$=1 supersymmetric Ising model, which has been  pointed out  in \cite{Shimada2016,Klebanov2016,Bashkirov2013}.

\vskip 2mm
\section{Conformal bootstrap for branched polymer}
\vskip 2mm

 We now consider the determinant method for the branched polymer. This determinant method will be applied to the RFIM in the next section.
  

The minor $d_{123}$ is defined by

 \be
d_{123}= {\rm det} \left(\begin{array}{ccc}
vs1 & vs2 & vs3 \\
vs1' & vs2' & vs3'\\
vt1 & vt2 & vt3 \\
\end{array} \right) 
\ee
where $vsn$ = $vsn(D,\Delta_\phi,\Delta_\epsilon)$ (n=1,2,3). The number $n$ is related to the derivative of the conformal block. The notation of vsn can be found in \cite{Hikami2017a}. $vsn'$ is a function of  $D, \Delta_\phi$ and $\Delta_1$.  $\Delta_1$ represents scalar scale dimension $\Delta_T$, which appears in the polymer case \cite{Shimada2016,Hikami2017b}.

 \vskip 4mm
{\bf Table 2} : The branched polymer in D dimension ; $\Delta_\phi= 3(D-3)/5$ (approximation)  and (A)the scale dimension of Yang-Lee model  in $D'= D-2$ dimension obtained from (\ref{brYL1}) ,
(B)the scale dimension of Yang-Lee model  in $D'=D-2$ dimension obtained from Pad\'e or exact solution(*) .\\
\\
\begin{tabular}[t]{|c|c|c|c|c|}
\hline
D & $\Delta_\phi$ (branched & $D'$&(A)  $\Delta_\phi$  (Yang-Lee & (B) $\Delta_\phi$  (Yang-Lee \\
 &   polymer)=3(D-3)/5 & =D-2 &edge singularity) & edge singularity)\\ 
\hline
D=3.0 & 0.0 & $D'$=1.0 & -1.0 & -1.0*\\
D=4.0 & 0.6  & $D'$=2.0   & - 0.4 & -0.4*\\
D=5.0 & 1.2 & $D'$=3.0 & 0.2  & 0.23\\
D=6.0 & 1.8 & $D'$=4.0 & 0.8 & 0.83\\
D=7.0 & 2.4  & $D'$=5.0   & 1.4  & 1.43\\
D=8.0 & 3.0 & $D'$=6.0  & 2.0 &  2.0*\\
\hline
\end{tabular}

\newpage
\vskip 3mm

For a polymer, which is represented in the limit $N\to 0$ in O(N) vector model, the conformal bootstrap method was  tapplied with 
the O(N) symmetric tensor scale dimension $\Delta_T$, which becomes equal to  $\Delta_\epsilon$ \cite{Hikami2017b}.
The O(N) symmetric tensor field  $\phi_{ab}(x)$ is given by
\be
\phi_{ab}(x) = :\phi_a\phi_b: -\frac{\delta_{ab}}{N} \sum_{m=1}^N :\phi_m^2:
\ee
and the energy density $\epsilon(x)$ is defined by
\be
\epsilon (x) = \sum_{m=1}^N : \phi_m^2 :.
\ee

The crossover exponent of O(N) vector model $\hat\varphi_2$ is given as
\be
\hat\varphi= \frac{D-\Delta_T}{D- \Delta_\epsilon}
\ee
and for a polymer (N=0), $\hat\varphi$ becomes 1, and it leads to the degeneracy of $\Delta_T=\Delta_\epsilon$ \cite{Shimada2016}.
The determinant method for the polymer with the scaling dimension $\Delta_T$ provides  good numerical values for the critical exponents \cite{Hikami2017b}.

For a branched polymer, which is represented by bosonic hamiltonian in (\ref{branch}), O(N) symmetric tensor scale dimension $\Delta_T$ is also important.
This scale dimension is a scalar, and we denote this for the branched polymer by $\Delta_1$ in the following.

 We put $\Delta_\phi= 3(D-3)/5$ as an approximation value for the branched polymer, which is not so different from the expected value in Tale 3, and we determine the value of $\Delta_\epsilon$ and $\Delta_1$ ($\Delta_1=\Delta_T$) from the intersection of
the zero loci of $3\times 3$ minors $d_{ijk}$. In Fig.3, we consider $D=8$. The intersection point shows $\Delta_\epsilon = \Delta_1= 4$. In Fig.4, $D=6$ is shown. In Fig.5, $D=4$ case
is shown with $\Delta_\phi=0.6$. The obtained value  $\Delta_\epsilon = 1.6$ is consistent with Yang-Lee edge singularity at D=2, $\Delta_\phi=\Delta_\epsilon = -0.4$.
Thus we find that, as in Fig. 3 - Fig.5, the dimensional reduction to $D'=D-2$ dimensional  Yang-Lee edge singularity holds.
\begin{figure}
{\bf Fig.3}\hskip 20mm {$\Delta_T$}\\
\centerline{\includegraphics[width=0.5\textwidth]{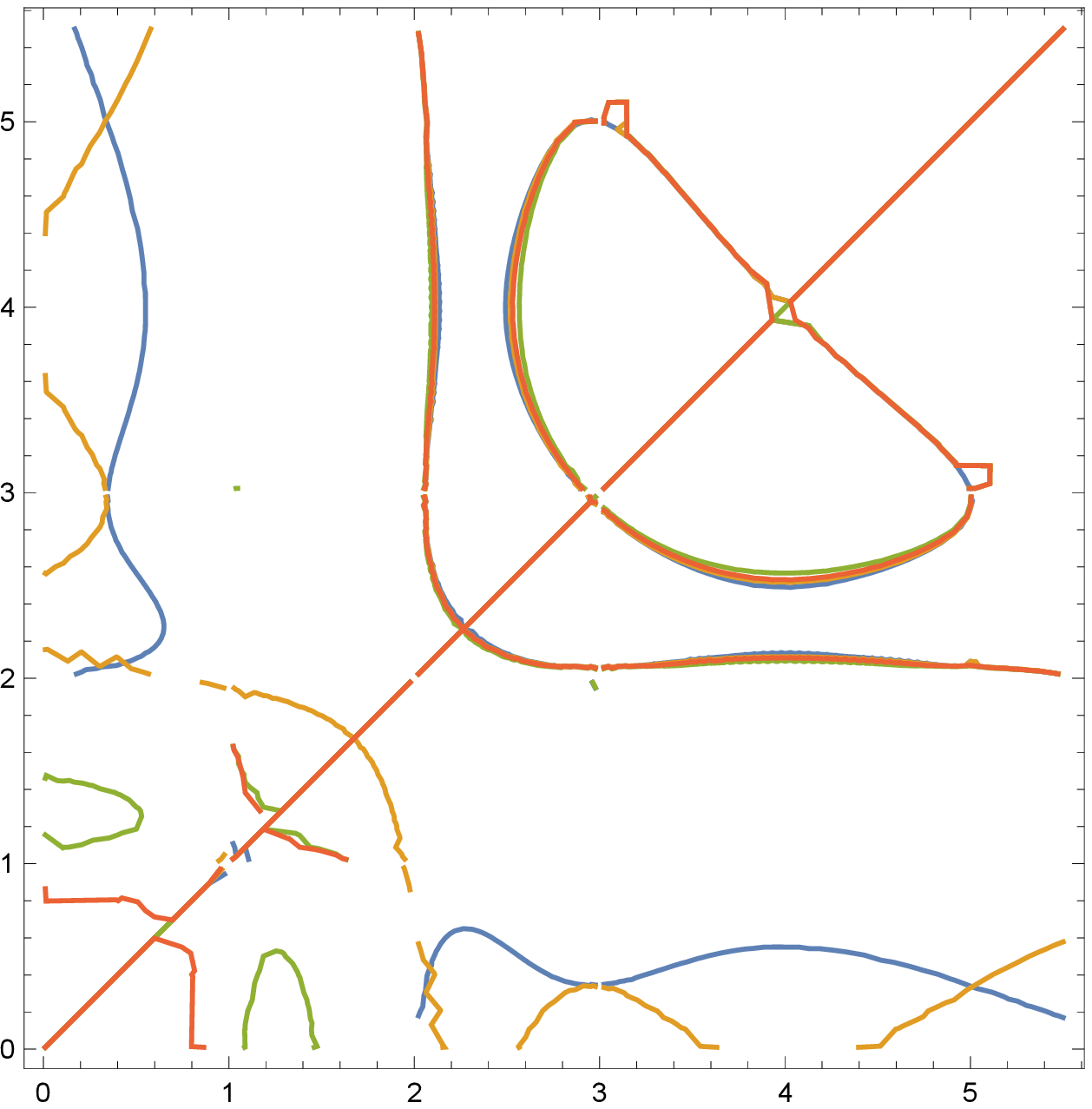}}\\
\hskip 90mm {$\Delta_\epsilon$}\\
\caption{ Branched polymer of D=8.0, $\Delta_\phi=2.0$: the contour of zero loci of $3\times3$ minors $d_{124},d_{123},d_{234},d_{134}$  are shown.  At $\Delta_\epsilon = 4.0$, the fixed point appears for a branched polymer. }
\end{figure}

\begin{figure}
{\bf Fig.4}\hskip 20mm {$\Delta_1$}\\
 \centerline{\includegraphics[width=0.5\textwidth]{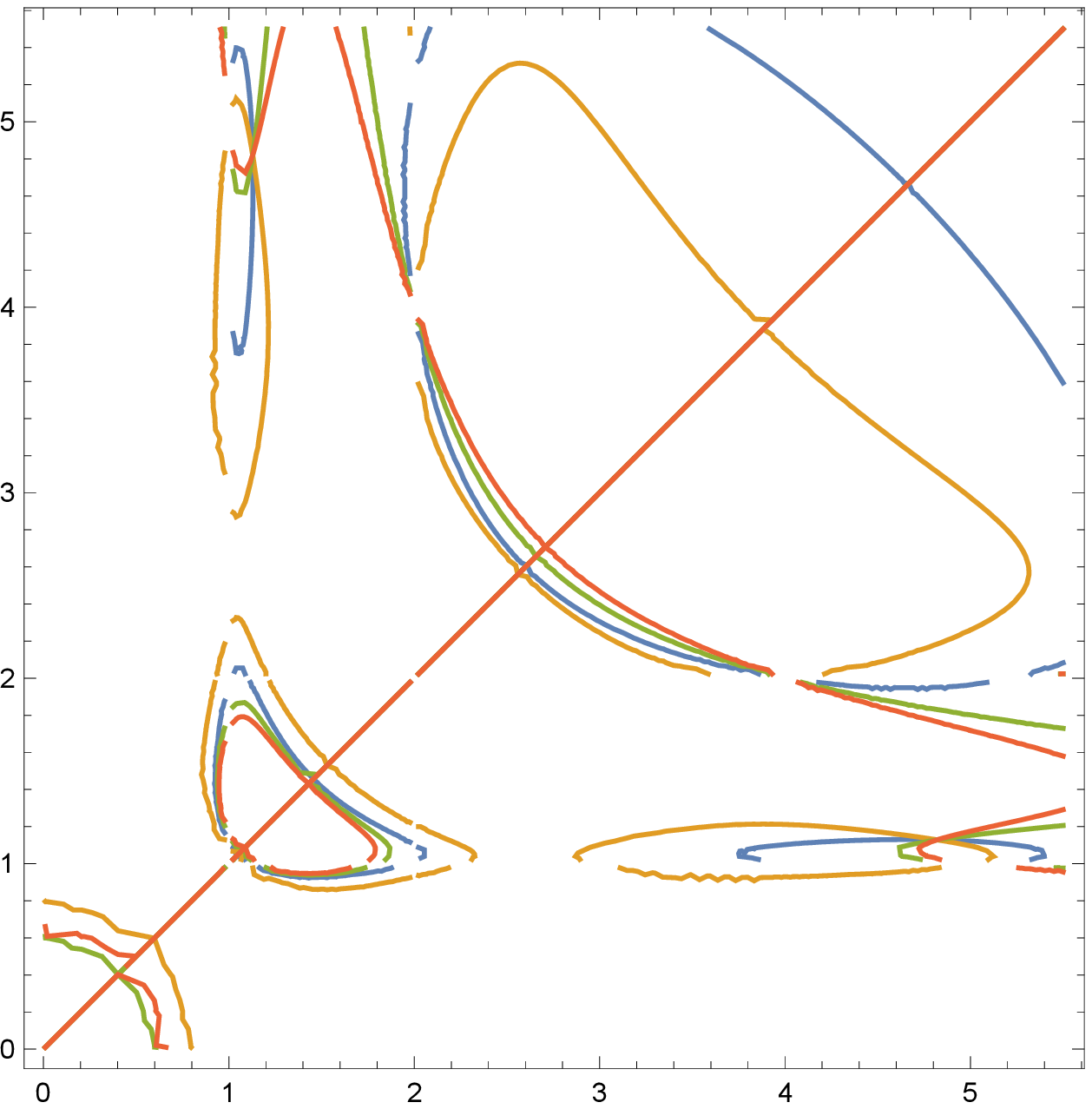}}\\
 \hskip 90mm {$\Delta_\epsilon$}\\
\caption{ Branched polymer of D=6; Four lines, the contour of zero loci of $d_{124},d_{123},d_{234},d_{134}$,  intersect with the line $\Delta_\epsilon =\Delta_1 (=\Delta_T)$ at $\Delta_\epsilon = 2.8$, which corresponds to Yang-Lee model at $D=4$ in Table 1. In this figure, $\Delta_\phi=0.83$ is taken.}
\end{figure}
\vskip 5mm

\begin{figure}
{\bf Fig.5}\hskip 20mm {$\Delta_1$}\\
\centerline{\includegraphics[width=0.5\textwidth]{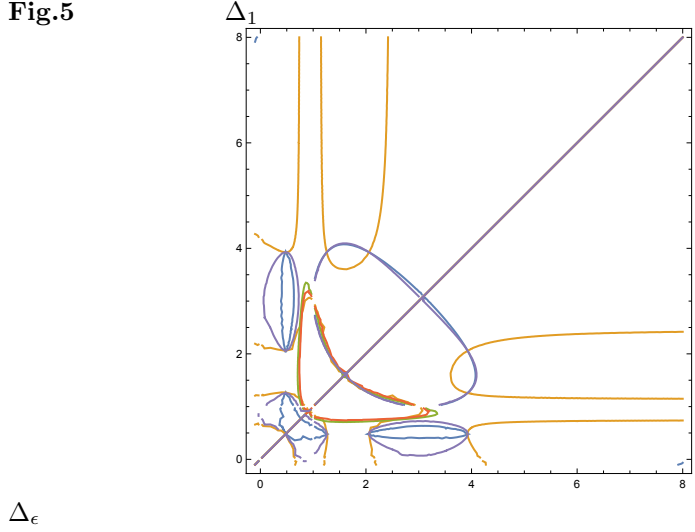}}\\
\hskip 90mm {$\Delta_\epsilon$}\\
\caption{ Branched polymer of D=4;  $3\times 3$ determinant method with $\Delta_\phi=0.6$: the contour of zero loci of $d_{124},d_{123},d_{234},d_{134},d_{125}$  are shown.  The fixed point is obtained at $\Delta_\epsilon=\Delta_1=1.6$ from the intersection of 5 lines of the zero loci. The correspondence of
$\Delta_\epsilon$=1.6 - 2 = - 0.4, to exact value of Yang-Lee edge singularity at D=2 is verified. The axis is $(x,y)= (\Delta_\epsilon,\Delta_1)$. This figure shows the blow up of the
singularity of $\Delta_\epsilon = \Delta_1$ to $\Delta_\epsilon\ne \Delta_1$. There is another fixed point at $\Delta_1=3.3$, which is considered as the correction to scaling $\Delta'_\epsilon$.}
\end{figure}
\vskip 5mm





\vskip 5mm

\vskip 2mm



\newpage
For $4\times 4$ minor method, the four scale dimensions are  $\Delta_\phi$, $\Delta_\epsilon$, $\Delta_1$ and $Q$. $Q$ is the spin 4 scale dimension.   For a polymer case, $\Delta_T= \Delta_\epsilon$, but for 
Ising model $\Delta_T$ is not equal to $\Delta_\epsilon$.  It takes a value near $\Delta_\epsilon$ \cite{Shimada2016}.

\vskip 2mm
{\bf D=6}
\vskip 2mm
 The fixed point at $\Delta_\phi=1.8$ and $\Delta_\epsilon=2.8$   is obtained for $D=6$. These scale dimensions are consistent with the dimensional reduction to $D'=4$ ($D'=D-2$) dimensional Yang-Lee edge singularity. The figure of this case has been represented in Fig.3 of \cite{Hikami2017b}.


\vskip 2mm
\vskip 2mm
{\bf D=5}
\vskip 2mm
For D=5, we find in Fig.6, a fixed of $\Delta_\phi=1.25,\Delta_\epsilon=2.4$ for the branched polymer. These scale dimensions are consistent with the dimensional reduction to Yang-Lee edge singularity of $D'=3$ dimensions ($D'=D-2$). We used the parameters of $Q=7.0,\Delta_1=2.6$ in Fig.6. 

\vskip 2mm



 \begin{figure}
 {\bf Fig.6} \hskip 20mm {$\Delta_\epsilon$}\\
 \centerline{\includegraphics[width=0.5\textwidth]{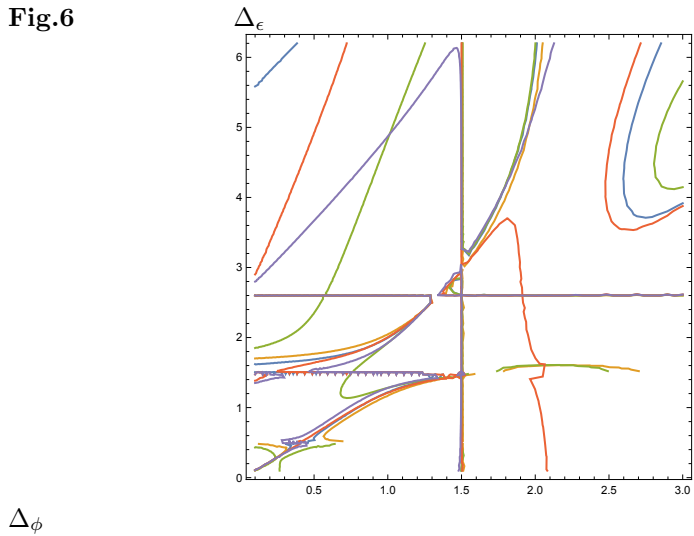}}\\
 \hskip 90mm {$\Delta_\phi$}\\
\caption{ Branched polymer in D=5 : the contour of  zero loci of minors are shown.   The fixed point $\Delta_\epsilon$ =2.45, $\Delta_\phi=1.25$ is obtained with $\Delta_1=2.6,Q=7.0$. This corresponds to $D'=3$ Yang-Lee edge singularity $\Delta_\epsilon=\Delta_\phi=0.23$ 
in Table 2. }
\end{figure}

\newpage

\section{Conformal bootstrap  for RFIM }
\vskip 2mm

 The relations between the scale dimension $\Delta_\phi$ and $\Delta_\epsilon$ of RFIM and pure Ising model are, if they hold
 \be\label{dr(1)}
 \Delta_\phi({\rm RFIM \hskip 1mm in \hskip 1mm D \hskip1mm dim.}) = \Delta_\phi({\rm pure\hskip 1mm Ising \hskip 1mm in \hskip 1mm D' \hskip 1mm dim.}) + 1
 \ee
 and
 \be\label{dr(2)}
 \Delta_\epsilon ({\rm RFIM \hskip 1mm in \hskip 1mm D\hskip 1mm dim.})= \Delta_\epsilon({\rm pure\hskip 1mm Ising  \hskip 1mm in \hskip 1mm D' \hskip 1mm dim. }) + 2
 \ee
where $D' = D - 2$.
 
 There are several arguments which explain the failure of aove dimensional reduction of RFIM. The most serious argument against the dimensional reduction may be
 the existence of the attractive potential of replica fields, which leads to the bound states \cite{Brezin2000,Parisi2002}. Recently, the break down of the dimensional reduction is suggested near $D=5$ \cite{Tarjus2011}. The recent review of RFIM may be found in \cite{Dotsenko2007,Sourlas2017}.
 
 Assuming that this dimensional reduction works near 6 dimensions for the random field Ising model, the conformal bootstrap method may be applied numerically for RFIM. From the dimensional reduction, we expect the correspondence of Table.3.
 \vskip 2mm
{\bf Table 3: Expected correspondence for RFIM to $D'=D-2$ dimensional Ising model} \\
 \begin{tabular}[t]{|c|c|c|c|c|c|}
\hline
D & $\Delta_\phi$ (RFIM) & $\Delta_\epsilon$ (RFIM) & $D'=D-2$ &  $\Delta_\phi$  (Ising) &  $\Delta_\epsilon$  (Ising) \\ 
\hline
D=6 & 2.0 & 4.0 & $D'=4$ & 1.0 & 2.0\\
D=5 & 1.516  & 3.414& $D'=3$   & 0.516 & 1.414\\
D=4 & 1.125 & 3.0 & $D'=2$ & 0.125  & 1.0\\
\hline
\end{tabular}
\vskip 6mm

{\bf Table 4. The scaling dimensions $\Delta_\epsilon$ and $\Delta_\phi$}: The  values $\Delta_\epsilon$ and $\Delta_\phi$ are obtained by the $4\times 4$ minors for values of $\Delta_1$ and $Q$. These results are obtained from the analysis in
Fig. 7 - Fig.13. The last column is $\epsilon$ expansion is [1,1] Pad\'e  up to the second order of $\epsilon$ ( $\Delta_\epsilon= 2- 2\epsilon/3+ 19 \epsilon^2/162 = (2-17/54 \epsilon)/(1+19/108 \epsilon)$). The value * is from\cite{ElShowk2014}, and the value** is exact value of 2D Ising model. The values of $\Delta_\epsilon$ in the fourth column agree well with the values of the sixth column for pure Ising model in $D'=D-2$ dimensions, when $D \ge 5$.
\vskip 2mm
\begin{tabular}[t]{|c|c|c|c|c|c|}
\hline
D & $\Delta_1$  & Q &$\Delta_\epsilon$ & $\Delta_\phi$  &  $\Delta_\epsilon$  (pure Ising in $D'$ dim.)+ 2 \\ 
\hline
D=6.0 & 4.3 &8.0&4.0  & 2.0 &4.0 \\
D=5.9 & 4.21  &7.9& 3.933   & 1.94997 &3.9345 \\
D=5.8 & 4.11 &7.8& 3.864 & 1.89996 & 3.8712  \\
D=5.5 & 3.81 & 7.5 &3.647 & 1.74995 & 3.6936 \\
D=5.0 & 3.18  & 7.0 & 3.41   &  1.49994 & 3.4331 (3.41267*)\\
D=4.5 & 2.7 & 6.5  & 2.93 & 1.25 &  3.2088\\
D=4.2 & 2.4 & 6.2 & 2.55 & 1.10 & 3.0886\\
D=4.0 &  2.2  &  6.0  &  2.0  & 1.0  & 3.0137(3.0**)\\
\hline
\end{tabular}
\newpage
\vskip 10mm

In Fig.7 and D=6, $\Delta_1=4.9$, we have a single fixed point at $\Delta_\phi=2,\Delta_\epsilon = 4$, which agrees with  D=4 Ising fixed point by the dimensional reduction. We examine the fixed points around $D=6$.
By another analysis, where a parameter $\Delta_1=4.3$ is chosen, we have  the same result. These values  exactly correspond to D=4 pure Ising model at D=4. Namely, the dimensional reduction of RFIM is valid for D=6. We note the value of 
$\Delta1$ is slightly different from $\Delta_\epsilon$.

\vskip 2mm
In Fig.8, D=5.9 case with $\Delta_1=4.21,Q=7.9$ is shown in the contour of the zero loci of 5 minors. The fixed point is located at $\Delta_\epsilon=3.933,\Delta_\phi=1.94997$,
which corresponds to D=3.9 pure Ising model by the dimensional reduction. In this Fig.10, there is a Gaussian fixed point at $\Delta_\epsilon=D-2 = 3.9$, which is infrared unstable.
The value of $\Delta_\phi$ is almost same as $(D-2)/2$, but slightly less than this value. This means that the exponent $\eta$ is negative.
\vskip 3mm
\begin{figure}
{\bf Fig.7}\hskip 20mm {$\Delta_\epsilon$}\\
 \centerline{\includegraphics[width=0.5\textwidth]{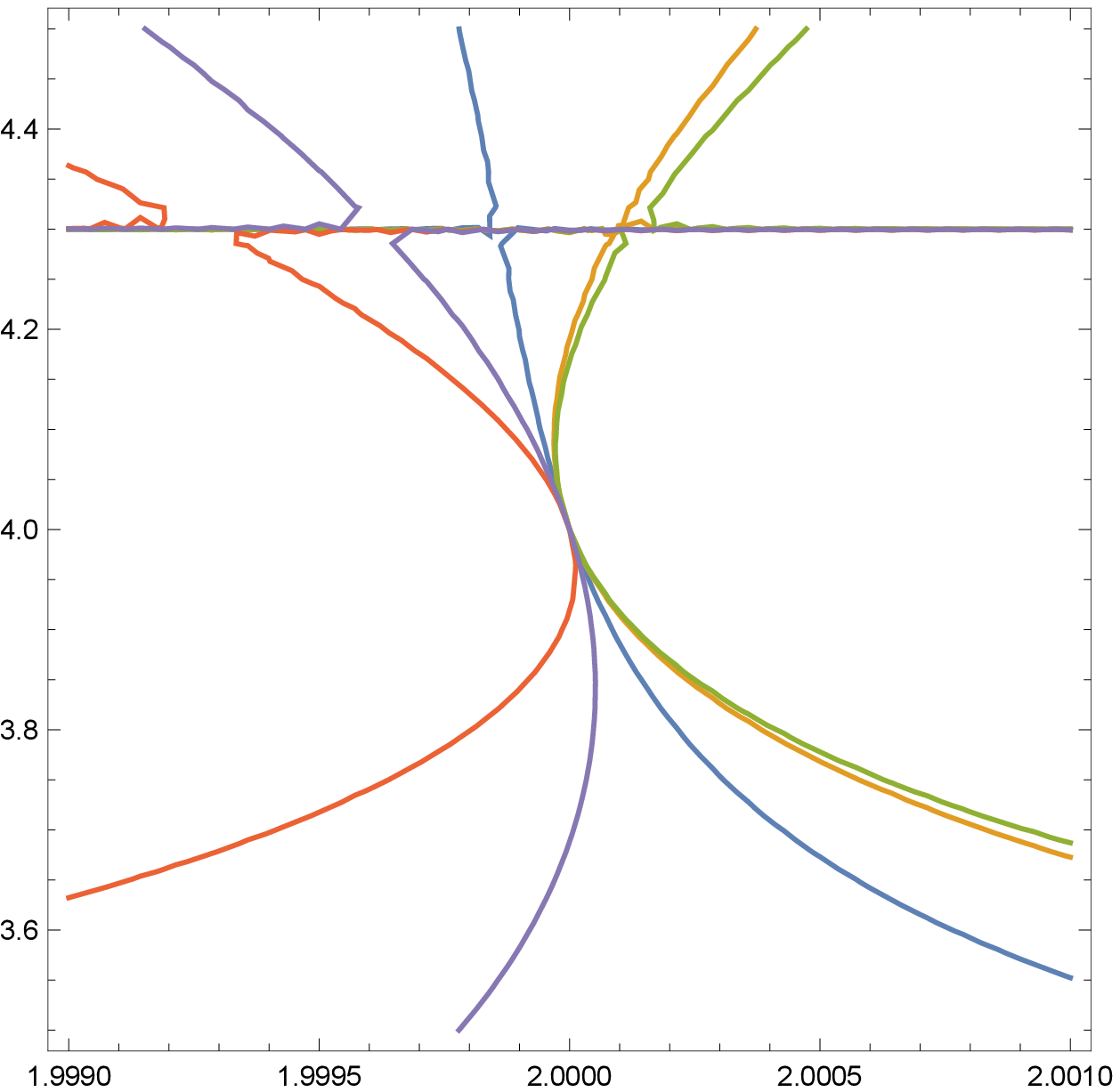}}\\
 \hskip 90 mm {$\Delta_\phi$}\\
\caption{ RFIM in D=6.0 : the 5 contours of  zero loci of  $4\times 4$ minors $d_{1234},d_{1245},d_{2345},d_{1245},d_{1235}$ are shown with $Q=8.0,\Delta_1=4.3$.   The fixed point $\Delta_\epsilon$ =4.0, $\Delta_\phi=2.0$ is obtained. }
\end{figure}

\begin{figure}
{\bf Fig.8} \hskip 20mm {$\Delta_\epsilon$}\\
 \centerline{\includegraphics[width=0.5\textwidth]{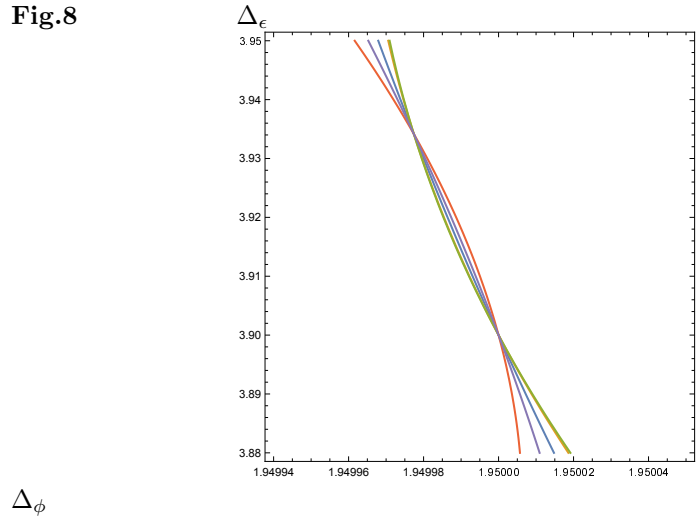}}\\
 \hskip 90mm {$\Delta_\phi$}\\
 \caption{ D=5.9 : the contour of  zero loci minors are shown with $Q=7.9,\Delta_1=4.21$.   The fixed point $\Delta_\epsilon$ =3.933, $\Delta_\phi=1.94997$ is obtained.  }
\end{figure}
\vskip 3mm

In Fig.9, D=5.8 case with $\Delta_1=4.11,Q=7.8$ is shown. The obtained values are $\Delta_\epsilon = 3.864,\Delta_\phi=1.89996$. This agrees with the dimensional reduction of pure Ising model at $D=3.8$ by the $\epsilon$ expansion, which gives $\Delta_\epsilon= 1.8645$.

\begin{figure}
{\bf Fig.9}\hskip 20mm {$\Delta_\epsilon$}\\
 \centerline{\includegraphics[width=0.5\textwidth]{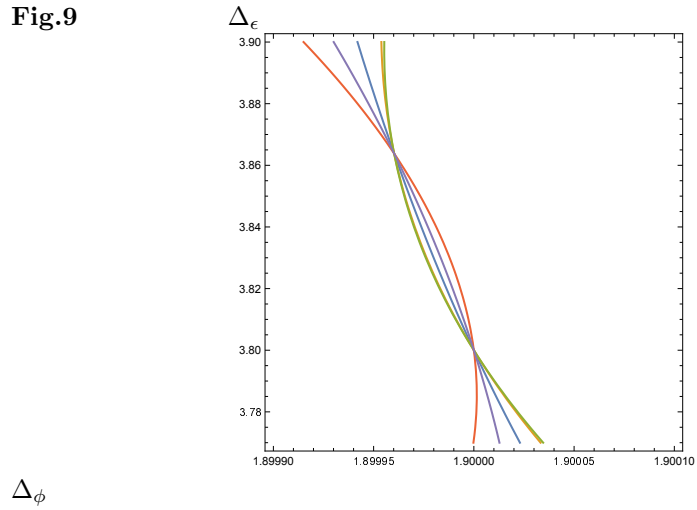}}\\
 \hskip 90 mm {$\Delta_\phi$}\\
\caption{ D=5.8 : the contour of  zero loci minors are shown with $Q=7.8,\Delta_1=4.11$.   The fixed point $\Delta_\epsilon$ =3.864, $\Delta_\phi=1.89996$ is obtained. }
\end{figure}
\vskip 3mm



\vskip 2mm

In  Fig.10, D=5.0 case with $\Delta_1=3.18,Q=7.0$ is shown in the contour of the zero loci of 5 minors. The fixed point at $\Delta_\epsilon =3.41,\Delta_\phi= 1.49994$ is obtained. This corresponds to pure D=3.0 Ising model($\Delta_\epsilon=1.414,\Delta_\phi=0.516$). The value of $\Delta_\epsilon$ of  pure D=3 Ising model
is 1.414, therefore $\Delta_\epsilon= 3.41$ agrees with the dimensional reduction, but the value of $\Delta_\phi$ disagrees. The value of $\Delta_\phi=1.49994$ corresponds to
$\eta/2=-0.00006$. If the value of $Q$ is changed to 7.04, the loci of minors do not intersect in a point, although the value of $\Delta_\phi$ approaches to the dimensional reduction of pure Ising model $\Delta_\phi=1.514$.

\begin{figure}
{\bf Fig. 10}\hskip 18mm {$\Delta_\epsilon$}\\
 \centerline{\includegraphics[width=0.5\textwidth]{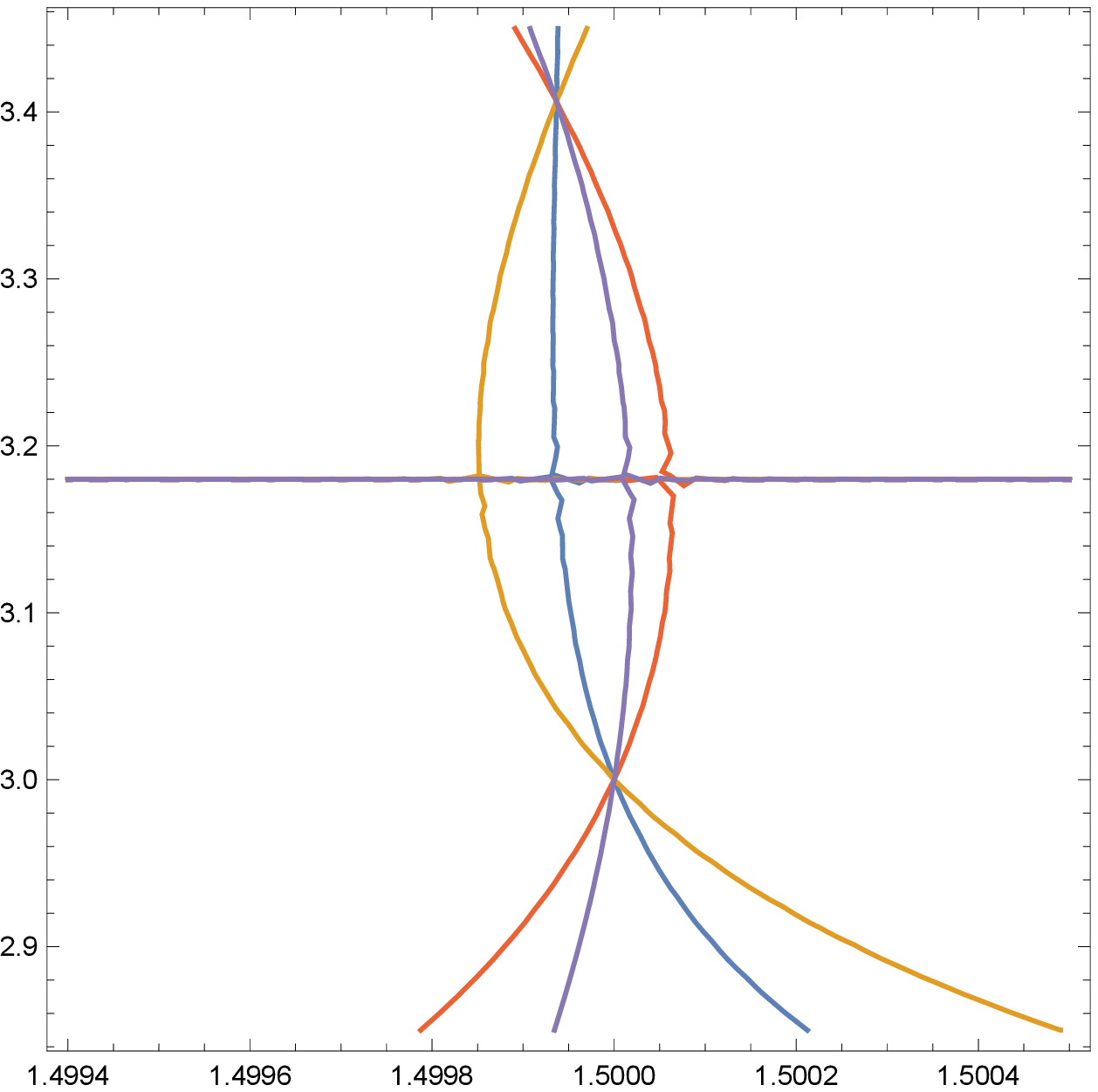}}\\
 \hskip 90 mm {$\Delta_\phi$}
\caption{ D=5.0 : the contour of  zero loci minors are shown with $Q=7.0,\Delta_1=3.18$.   The fixed point $\Delta_\epsilon$ =3.41, $\Delta_\phi=1.49994$ is obtained. }
\end{figure}

For D=4.5 case with $\Delta_1=2.7,Q=6.5$,  the fixed point at $\Delta_\epsilon=2.93,\Delta_\phi=1.2502$ is obtained.
\vskip 2mm
For D=4.2 case with $\Delta_1=2.4,Q=6.2$, the fixed point is located at $\Delta_\phi=1.10$ and $\Delta_\epsilon= 2.55$. As shown in Table 4, the deviation of the value of $\Delta_\epsilon$ is large from the expected value by Pad\'e value, which is $ 3.0886$.

\vskip 2mm
In Fig.11, D=4.1 case is shown with $\Delta_1=2.3,Q=6.1$. The fixed point $\Delta_\epsilon= 2.3, \Delta_\phi=1.05$ is obtained.

\vskip 2mm
In Fig.12 and 13, D=4.0 case is shown with $\Delta_1= 2.2, Q=6.0$. The fixed point $\Delta_\epsilon=2.0,\Delta_1.0$ is obtained , it is Gaussian fixed point. Fig.13 is a global map.
It is remarkable that we obtain the free field fixed point at D=4. This is due the small value of $\Delta_1$. When we take large value of $\Delta_1$, there appears ordinary Ising fixed point. Indeed when D=3.9, as shown in Fig.19, for the larger value of $\Delta_1=4.0$, $Q=5.9$, we obtain an Ising fixed point at $\Delta_\epsilon = 1.92666$, $\Delta_\phi=0.95003$.

\begin{figure}
{\bf Fig.11}\hskip 18mm {$\Delta_\epsilon$}\\
\centerline{\includegraphics[width=0.5\textwidth]{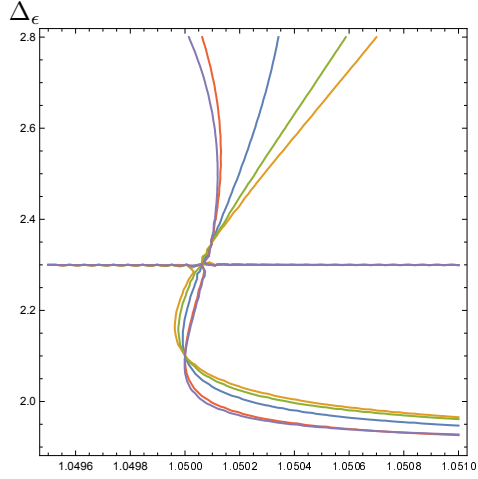}}\\
\hskip 90mm {$\Delta_\phi$}\\
\caption{ D=4.1 : the contour of  zero loci of minors are shown with $Q=6.1,\Delta_1=2.3$.   The fixed point $\Delta_\epsilon$ =2.3, $\Delta_\phi=1.05$ is obtained. }
\end{figure}

\begin{figure}
{\bf Fig.12}\hskip 18mm {$\Delta_\epsilon$}\\
 \centerline{\includegraphics[width=0.5\textwidth]{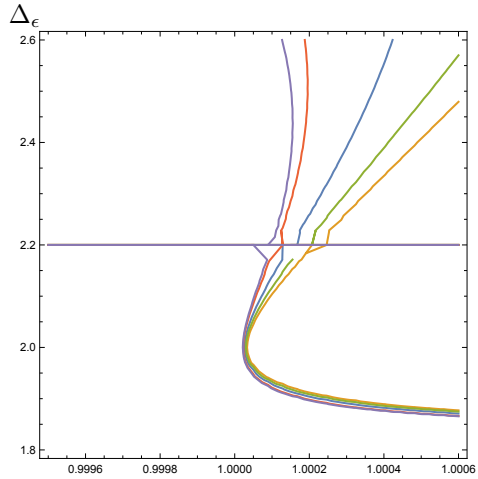}}\\
 \hskip 90mm {$\Delta_\phi$}\\
\caption{ D=4.0 : the contour of  zero loci of minors are shown with $Q=6.0,\Delta_1=2.2$.   The fixed point (Gaussian) $\Delta_\epsilon$ =2.0, $\Delta_\phi=1.0$ is obtained. }
\end{figure}

\begin{figure}
{\bf Fig.13}\hskip 18mm{$\Delta_\epsilon$}\\
 \centerline{\includegraphics[width=0.5\textwidth]{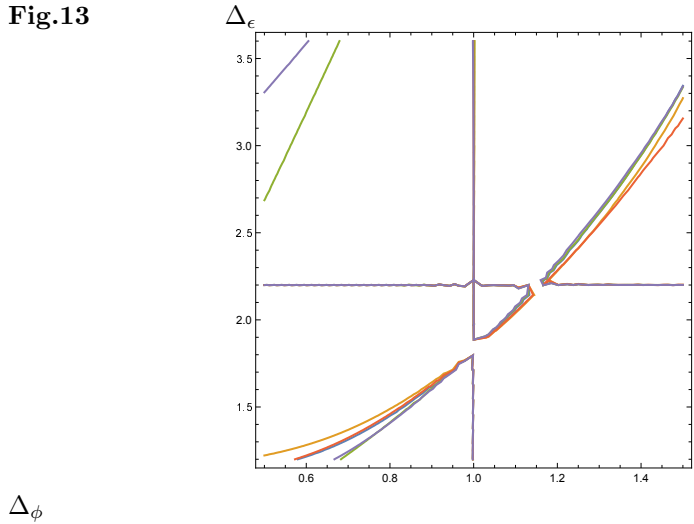}}\\
 \hskip 90mm {$\Delta_\phi$}\\
\caption{ D=4.0 (global map): the contour of  zero loci of minors are shown with $Q=6.0,\Delta_1=2.2$.   The fixed point (Gaussian) $\Delta_\epsilon$ =2.0, $\Delta_\phi=1.0$ in Fig.17 is on the line $\Delta_\phi=1.0$. }
\end{figure}

  \newpage
 \vskip 3mm
 \section{Summary and discussions}
\vskip 2mm

We have discussed in this paper, the conjectures of the dimensional reductions of  the branched polymer to Yang Lee edge singularity  and RFIM to the pure Ising model by the small $4\times 4$ minors.
The conformal bootstrap determinant method  gives the confirmations of the dimensional reductions in the branched polymer within the numerical  approximations. The critical dimension of a branched polymer
 is 8, and it corresponds to $D'=6$ dimensional Yang-Lee edge singularity.
We have confirmed for $4< D<8$, there is a fixed point which is $\Delta_\epsilon= \Delta_\epsilon$ (Yang-Lee in $D'=D-2$ dimensions)+2, $\Delta_\phi=\Delta_{\phi}$(Yang-Lee in $D'= D-2$ dimensions) + 1. With the relation of Yang-Lee edge singularity model $\Delta_\epsilon = \Delta_\phi$, we obtain 
\be
\Delta_\epsilon = \Delta_\phi + 1
\ee
which is a relation appeared  for  $\mathcal{N}$= 1 supersymmetric Ising model \cite{Bashkirov2013, Shimada2016, Klebanov2016}. These results are reported in the previous paper  \cite{Hikami2017b}.

For RFIM, the upper critical dimension is 6. For $D < 6$, there appears a fixed point, which agrees with the values of $\epsilon$ expansion of $\Delta_\epsilon$,
but the value of $\eta$ becomes negatively small for $D < 6$. The result is summarized in Table 4. For $5<D<6$, the values of $\Delta_\epsilon$ is almost consistent with the  $\epsilon$ expansion, with appropriate values of $Q$ and $\Delta_1$.
However, for $D<5$ the deviation becomes quite large, and the conjecture of the correspondence of the dimensional reduction is violated for $D<5$ as seen in Table 4.

The bound state has been suggested in the literatures \cite{Brezin2000,Parisi2002} for the explanation of this fail of the dimensional reduction. The peculiar almost straight line as shown in Fig. 8 and Fig.9 may indicate the bound state (also related to the negative small value of $\eta$). This finding may be  consistent with the formation of  the bound state.
The breakdown for $D<5$ seems to  be consistent with the recent results of \cite{Tarjus2011,Sourlas2017}. The recent works \cite{Sourlas2017} also shows the dimensional reduction of RFIM in $D=5$ to pure Ising model in $D'=3 $works precisely.  From the point of view of the supersymmetry, we observed the difference between Ising tri-critical point ($\phi^6$ theory) and $\mathcal{N}$=1 supersymmetric fixed point \cite{Shimada2016}, although it is well known that in two dimensions, the tri-critical point coincides with the supersymmetric point \cite{Friedan1984}.
It is interesting to investigate the relation of RFIM to the multi critical behaviors such as tri-critical Ising model for $D<5$. For such study, we need more scale dimensions 
of OPE in addition to the dimension $\Delta_1$ which is studied here.

The random field for O(N) vector model gives also the dimensional reduction as (\ref{ON}). We considered only RFIM, which is $N=1$ of $O(N)$ vector model. It may be important to
investigate the conformal bootstrap analysis for $O(N)$ vector spin model with random field model. This study will be a future work. As a disordered system, there is a problem of Anderson localization with the spin orbit interaction \cite{Hikami1980}, which has a phase transition in two dimensions. It is related to replica limit and the supersymmetry. We will discuss this problem in a coming article \cite{Hikami2019}.

 \vskip 5mm
{\bf Acknowledgements}
\vskip 3mm
Author is thankful to Ferdinando Gliozzi for the discussions of the determinant method.  He  thanks Edouard Br\'ezin
 for the discussions of the dimensional reduction problem in branched polymers and suggestion  of RFIM problem by conformal bootstrap method. Part of this work was presented in the workshop of BMFT in Rome University in January 3rd, 2018 and author thanks Georgi Parisi for this invitation. 
 This work is supported by JSPS KAKENHI Grant-in-Aid 16K05491 and 19H01813.
\newpage

 \newpage

\end{document}